\documentclass[a4paper,11pt]{article}
\usepackage{jinstpub} 
\usepackage{lineno} 

\title{Front-end electronics development of large-area SiPM arrays for high-precision single-photon time measurement}

\author[a]{W. Zhi}
\author[b]{, R.K. Cao}
\author[a]{, J.N. Tang}
\author[a]{, M.X. Wang}
\author[a]{, Y.Q. Tan}
\author[a,1]{, W.H. Wu\note{Corresponding author.}}
\author[a,b]{and D.L. Xu}

\affiliation[a]{School of Physics and Astronomy, Shanghai Jiao Tong University \\
                800 Dongchuan Rd, Minhang District, Shanghai, China}
\affiliation[b]{Tsung-Dao Lee Institute, Shanghai Jiao Tong Univeristy \\
                520 Shengrong Rd, Pudong New Area, Shanghai, China}

\emailAdd{wuweihao@sjtu.edu.cn}

\abstract{
    TRopIcal DEep-sea Neutrino Telescope (TRIDENT) plans to incorporate silicon photomultipliers (SiPMs) with superior time resolution in addition to photomultiplier tubes (PMTs) into its detection units, namely hybrid Digital Optical Modules (hDOMs), to improve its angular resolution.
    However, the time resolution significantly degrades for large-area SiPMs due to the large detector capacitance, posing significant challenges for the readout electronics of SiPMs in hDOM.
    We analyzed the influences of series and parallel connections when constructing a large-area SiPM array and designed a series-parallel connection SiPM array with differential output.
    We also designed a high-speed pre-amplifier based on transformers (MABA-007159) and radio frequency amplifiers (BGA2803), and an analog multi-channel summing circuit based on operational amplifiers (LMH6629).
    We measured the single photon time resolution (SPTR) of a $4\times4$ SiPM (Hamamatsu S13360-3050PE) array ($12\times12~\mathrm{mm}^2$) of approximately 300 ps FWHM.
    This front-end readout design enables the large-area SiPM array to achieve high-precision single photon time measurement in one readout channel.
}


\arxivnumber{2403.02948} 

\begin{document}
\maketitle
\flushbottom

\section{Introduction}
\label{sec1}

High-energy neutrinos from astrophysical sources can travel almost undisturbedly to Earth due to their small weak interaction cross section, providing direct information about their sources.
TRopIcal DEep-sea Neutrino Telescope (TRIDENT) is a next-generation neutrino telescope planned to be constructed in the South China Sea~\cite{TRIDENT_NA}.
As a prospective development direction, neutrino telescopes require improved angular resolution to more precisely detect and locate astrophysical neutrino sources~\cite{cosmic_ray_neutrino_telescope}.
Improving angular resolution can be achieved with superior time resolution for neutrino telescopes since they reconstruct neutrinos' information by detecting the Cherenkov light produced by secondary charged particles generated from weak interactions.
However, photomultiplier tubes (PMTs), commonly employed as photodetectors in neutrino telescopes, typically have a transit time (TT) on the order of tens of nanoseconds and the transit time spread (TTS) at the nanosecond level~\cite{IceCube_PMT}.
To enhance time resolution, TRIDENT is currently exploring the feasibility of incorporating fast-response silicon photomultipliers (SiPMs), into its detection units, namely hybrid Digital Optical Modules (hDOMs)~\cite{hDOM_ICRC2021, SiPM_ICRC2023}.
SiPM, with a sub-nanosecond single photon time resolution (SPTR), is a solid-state semiconductor photodetector composed of thousands of micro cells capable of detecting single photons~\cite{Hamamatsu_MPPC_technote, TOF_PET_SPTR}.
With its compact size, ease of integration, and low operating voltage, SiPM is increasingly being employed in the field of single-photon detection, such as time-of-flight positron emission tomography (TOF-PET) and calorimeters~\cite{SiPM_TOF_PET, SiPM_in_particle_physics}.

TRIDENT plans to construct about 1200 strings following a Penrose tiling distribution with a horizontal distance between two adjacent strings of 70 m or 110 m, and each string contains 20 hDOMs with a vertical distance between two hDOMs of 30 m~\cite{TRIDENT_NA}.
hDOM can only receive a few photons with the photon number mostly at the single-photon level because Cherenkov light has a low light yield and attenuates significantly in seawater over long distances.
To improve detection efficiency, the hDOM intends to comprise 24 SiPM arrays in addition to 31 PMTs, where PMTs have a large photon collection area and SiPM arrays exhibit high-precision time resolution~\cite{hDOM_ICRC2021}.
The SiPM array is aimed at achieving a detection area on the order of square centimeters and a SPTR at the order of hundreds of picoseconds, to enhance the effective area and time resolution of hDOM and improve the angular resolution of the neutrino telescope.
At the same time, due to the large number of hDOMs in TRIDENT, it is essential to minimize the power consumption and the number of readout channels of SiPM arrays.

However, the increased area will seriously affect the time resolution of SiPMs due to the larger detector capacitance, which increases the influence of electronic noise~\cite{Analysis_SPTR}.
For instance, a typical SPTR full width at half maximum (FWHM) result for a SiPM with a size of $1\times 1~ \mathrm{mm}^2$ is approximately 80 ps, whereas this value for a size of $3\times 3~ \mathrm{mm}^2$ can be around 180 ps~\cite{SPTR_SPAD_to_SiPM}.
To mitigate the effects of electronic noise, some front-end readout designs incorporate balun transformers and radio frequency (RF) amplifiers, achieving measured SPTRs of less than 150 ps FWHM for $3\times 3~ \mathrm{mm}^2$ Hamamatsu SiPMs and less than 100 ps FWHM for $4\times 4~ \mathrm{mm}^2$ FBK SiPMs, respectively~\cite{SiPM_improved_SPTR_measurement}.
Besides, some application-specific integrated circuits (ASICs), such as NINO and PETIROC, have also obtained SPTRs of about 200 ps FWHM for many $3\times 3~ \mathrm{mm}^2$ SiPMs~\cite{ASIC_NINO, SPTR_state, ASIC_Petiroc}.
However, achieving high-precision single-photon time measurements with these schemes will require a large number of readout channels for a SiPM array of square centimeter scale.

Focusing on high-precision time measurements based on large-area SiPM arrays, we analyze the existing challenges and propose an overall front-end readout scheme in section~\ref{sec2}. 
In section~\ref{sec3}, we propose the design of the front-end readout electronics for the SiPM array, including the series-parallel connection SiPM array and the two-stage amplification scheme.
Additionally, we present the experimental setup and some results and discussions of the SPTR measurements in section~\ref{sec4}.

\section{SiPM array front-end readout scheme}
\label{sec2}

As mentioned in section~\ref{sec1}, the primary challenge in the front-end readout of the SiPM array is maintaining its high-precision time measurement capability while constructing a large-area SiPM array in one readout channel.
Initially, the pre-amplifier requires a sufficiently high bandwidth to handle the high-speed leading edge signals from the SiPM, where the SiPM (Hamamatsu S13360-3050PE) exhibits a rapid rise time at around 1 ns~\cite{Datasheet_S13360, SiPM_understanding}.
The rise time and signal-to-noise ratio (SNR) of the waveform will directly influence the uncertainty of the time measurement results~\cite{Analysis_SPTR}.
Moreover, to increase the detection area on one channel, it is necessary to combine multiple SiPMs with a size of such as $3\times 3~ \mathrm{mm}^2$ to a pre-amplifier, because it is improper to configure one amplifier to each SiPM under the compact space and limited power consumption constraints of hDOM.
Constructing a large-area SiPM array can be achieved with a series and parallel connection~\cite{SiPM_array_construction}.

\begin{figure}[htbp]
    \centering
    \begin{minipage}[c]{0.32\textwidth}
        \centering
        \includegraphics[width=1\textwidth]{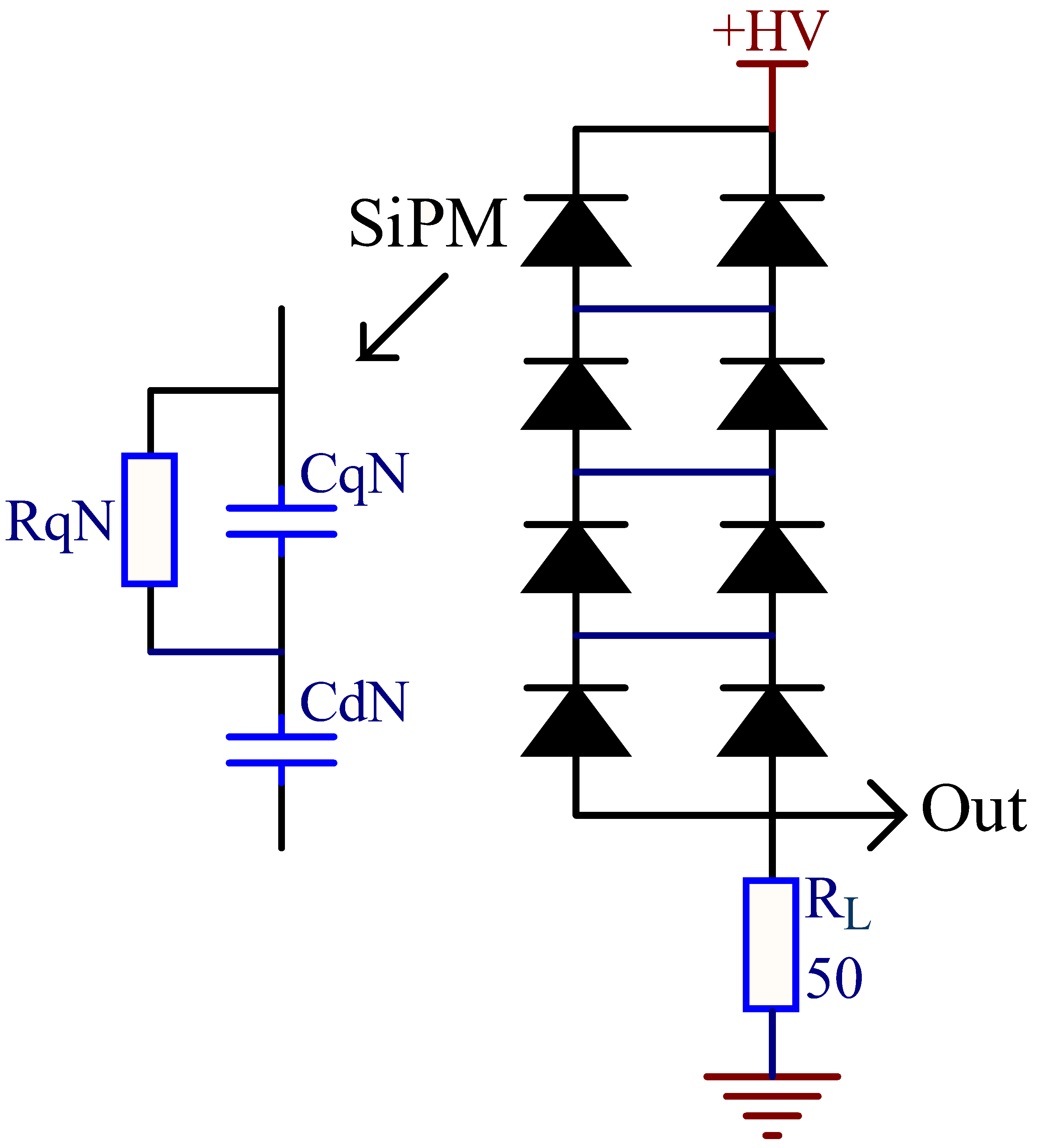}
      \end{minipage}
      \hfil
      \begin{minipage}[c]{0.6\textwidth}
        \centering
        \includegraphics[width=1\textwidth]{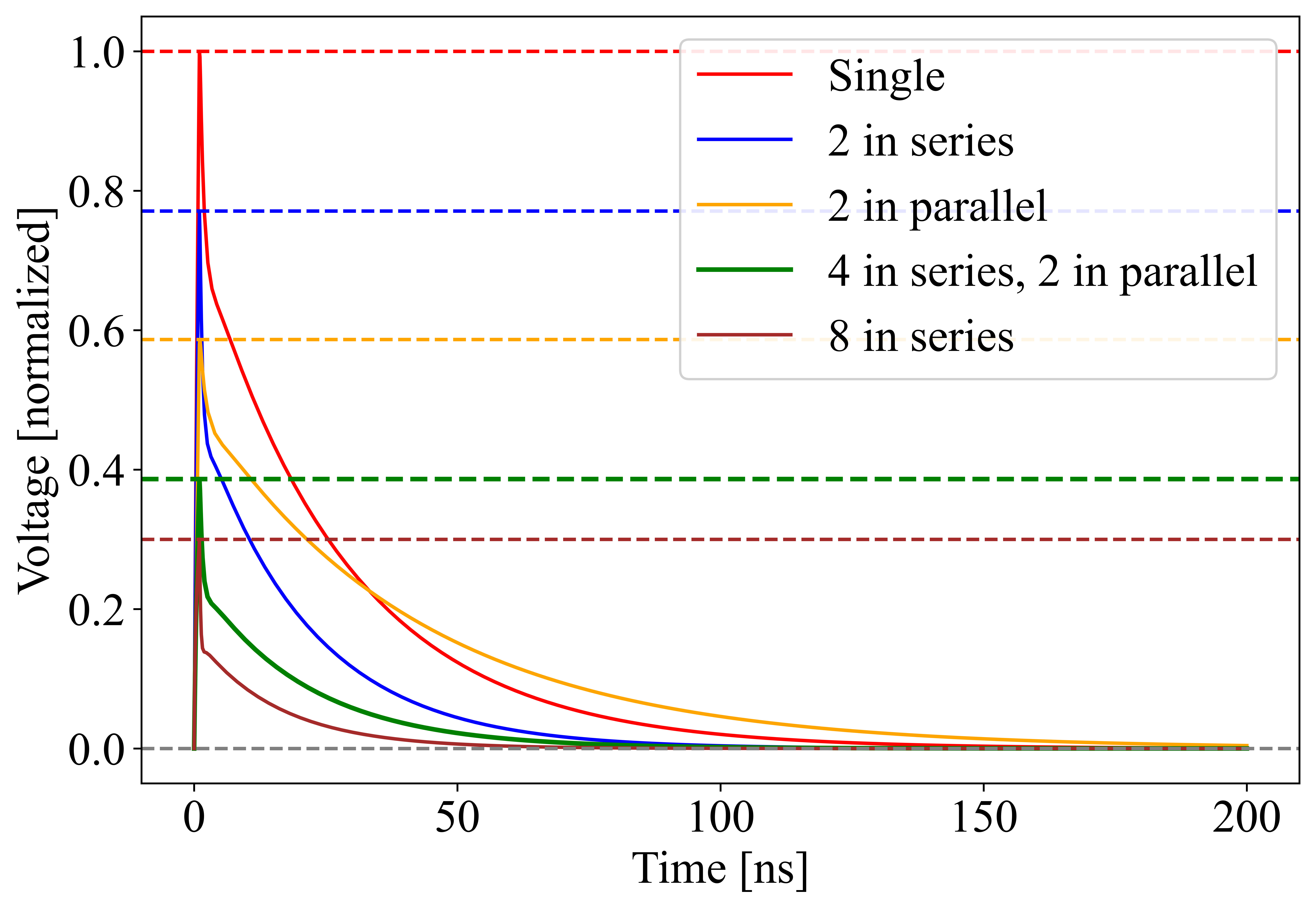}
      \end{minipage}
    \caption{Left: schematic of a series-parallel SiPM array utilized in simulations, where the SiPM model is simplified using resistors and capacitors.
             SiPM's parameters include $\mathrm{N=3600,~R_q = 160~k\Omega}$, $\mathrm{C_q = 7~fF}$, $\mathrm{C_d = 80~fF}$, and the avalanche process is simplified to a current pulse lasting a nanosecond~\cite{SiPM_parameters}.
             Right: simulation waveforms of SiPMs in various configurations, with the signal output through a $50~\Omega$ grounded resistor.
             The red solid line represents the waveform of a single SiPM, with all other waveforms normalized relative to this reference.
             The blue and orange solid lines represent waveforms of SiPM arrays with two SiPMs in different configurations, while the green and brown solid lines are waveforms of SiPM arrays with eight SiPMs.}
    \label{fig:2.simulation_para_series}
\end{figure}

In terms of electrical performance, the main impact of series and parallel connection is the reduction in signal amplitude, corresponding to the decrease in SNR.
Consider a series-parallel SiPM array and a readout circuit with an input impedance of $\mathrm{R_\text{L} = 50~\Omega}$, as shown in figure~\ref{fig:2.simulation_para_series}.
For the commonly used parallel connection scheme, the increased capacitance will decrease the signal's amplitude with a ratio of about $1/N_{\text{parallel}}$. 
This is also why directly choosing large-area SiPMs is not feasible, as SiPMs themselves are constructed with cells in parallel.
A simulation result based on Tina TI shows the influence on waveforms of series and parallel configurations, as shown in figure~\ref{fig:2.simulation_para_series}.
The signal amplitude in the series connection decreases with a ratio of approximately $\mathrm{R_L}/\left(N_{\text{series}}\cdot\mathrm{Z_{SiPM}}+\mathrm{R_L}\right)$ due to the SiPM impedance $\mathrm{Z_{SiPM}}\approx 1/\left(i~\omega \cdot \mathrm{C_{SiPM}}\right)\sim 10~\Omega$ on the series path, whereas this effect is much less than that of parallel connection. 
As a result, the series connection can be used as the main method to construct a series-parallel array, while parallel connections can be introduced to reduce impedance when the number of SiPMs in series is large.
For instance, in an array comprising eight SiPMs, a configuration with four SiPMs in series and two series in parallel is more feasible than having all eight SiPMs in series alone due to the higher impedance of the latter, as shown in figure~\ref{fig:2.simulation_para_series}.

Another factor influencing the timing performance is the difference in signal path length in series connections.
In a series, the pulsed signal from one SiPM must pass through other SiPMs, leading to fixed differences in path lengths from different SiPMs to the readout circuit.
The differences in signal path lengths among SiPMs in a series result in time deviations due to the finite speed of signal propagation, ultimately contributing to the total SPTR.
Assuming that electrical signals travel at about 2/3 of the speed of light on the printed circuit board (PCB), for a SiPM (S13360-3050PE) with an overall size of about 4 mm, this could result in a time deviation of $\text{20 ps}$ between the signals of two SiPMs in series.
To estimate this impact, consider an arithmetic sequence of number $N$ with a difference of 20 ps.
The standard deviation of this distribution is given by $\mathrm{\sqrt{(N^2-1)/12} \cdot 20~ps}$.
For instance, for an array with four SiPMs in series, this yields about 22 ps, contributing about $\text{53 ps}$ FWHM to SPTR.
Besides, it is worth mentioning that some devices, such as SiPMs, can introduce additional time delays because of their internal structure.

The number of SiPMs that can be combined into a series-parallel SiPM array is limited by the decrease in signal amplitude introduced by both series and parallel connections, as well as the signal path differences introduced in the series connection.
To further increase the number of SiPMs on one channel, multiple series-parallel channels can be integrated through an multi-channel analog summing circuit.
As a result, the front-end readout electronics design for the SiPM array includes two aspects detailed in section~\ref{sec3}:
the series-parallel connection SiPM array design and a two-stage amplification design.
The first stage is the pre-amplifier for a SiPM or series-parallel SiPM array, and the second stage is the multi-channel analog summing circuit.



\section{Front-end readout electronics design}
\label{sec3}

\subsection{Series-parallel connection SiPM array design}

\begin{figure}[htbp]
  \centering
  \includegraphics[width=0.5\textwidth]{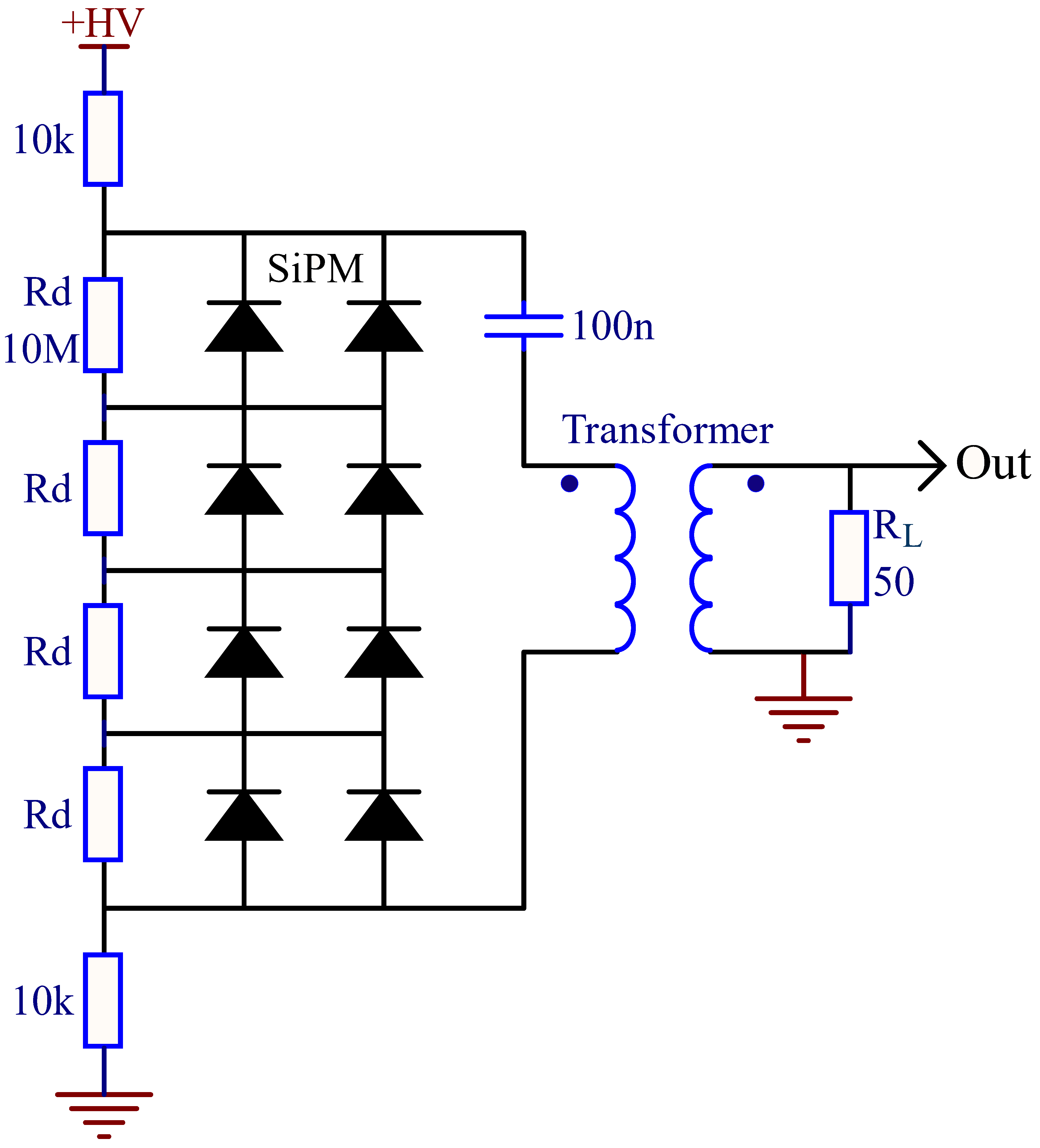}
  \caption{Schematic of series-parallel connection SiPM array, illustrating four SiPMs in series and two series in parallel.
           $\mathrm{R_d}$ is the voltage divider resistor with a typical value of 10 M$\Omega$, and $\mathrm{R_L}$ represents the input impedance of the pre-amplifier.}
  \label{fig:3.series_parallel_configuration}
\end{figure}

The series-parallel connection SiPM array utilizes a primarily series connection with supplementary parallel connections, enabling multiple SiPMs to be connected to a channel for output and reducing the influence of capacitance.
SiPMs located near the HV port exhibit longer signal paths in figure~\ref{fig:2.simulation_para_series}, leading to a significant time difference, which is a disadvantage of single-ended output configurations.
However, the differential output can help mitigate signal path discrepancies.
As a result, the series-parallel connection design scheme is illustrated in figure~\ref{fig:3.series_parallel_configuration}.
The transformer is used to realize the differential output and convert the differential signal into a single-ended one.
We conducted some experiments to showcase the improvements in signal path differences for differential output, as discussed in section~\ref{sec4}.

For a series-parallel SiPM array, the differences between SiPMs also affect the overall time performance~\cite{Analysis_SPTR}.
All SiPMs on one series-parallel SiPM array will receive the same bias voltage, but due to inherent variations in their breakdown voltages, there will be differences in overvoltage.
This can lead to variations in gain, introducing deviations in signal amplitude and resulting in time walk effects in time measurements.
Estimating its impact, consider a simple waveform in the linear form $U(t)=t\cdot V/T$, where $V$ is the peak value and $T$ is the time to rise from zero to the peak.
The trigger time at which the waveform reaches the threshold $V_{th}$ is given by $t = T \cdot V_{th}/V$.
If $V$ has a small variation $\Delta V$, the variation in trigger time is $\Delta t \approx T\cdot\frac{\Delta V}{V} \cdot \frac{V_{th}}{V}$.
Assuming $T=1ns$, $\frac{\Delta V}{V}=1/10$ and $\frac{V_{th}}{V}=1/2$, the calculated result yields $\Delta t \approx 50$ ps.
The standard deviation of breakdown voltages is about 0.4 V for 20,000 pieces of SiPMs (S13360-3050PE) we have purchased from Hamamatsu, which means that the $\frac{\Delta V}{V}$ is less than 1/10 when the average overvoltage exceeds 4 V.
Alleviating this effect can be achieved by selecting SiPMs with similar characteristics such as breakdown voltage and connecting them to the same series-parallel SiPM array, which can be conveniently done using the test data provided by the manufacturer.
Additionally, increasing the bias voltage or reducing the threshold voltage can also mitigate this impact.

\subsection{Two-stage amplification design}

\begin{figure}[htbp]
    \centering
    \includegraphics[width=0.72\textwidth]{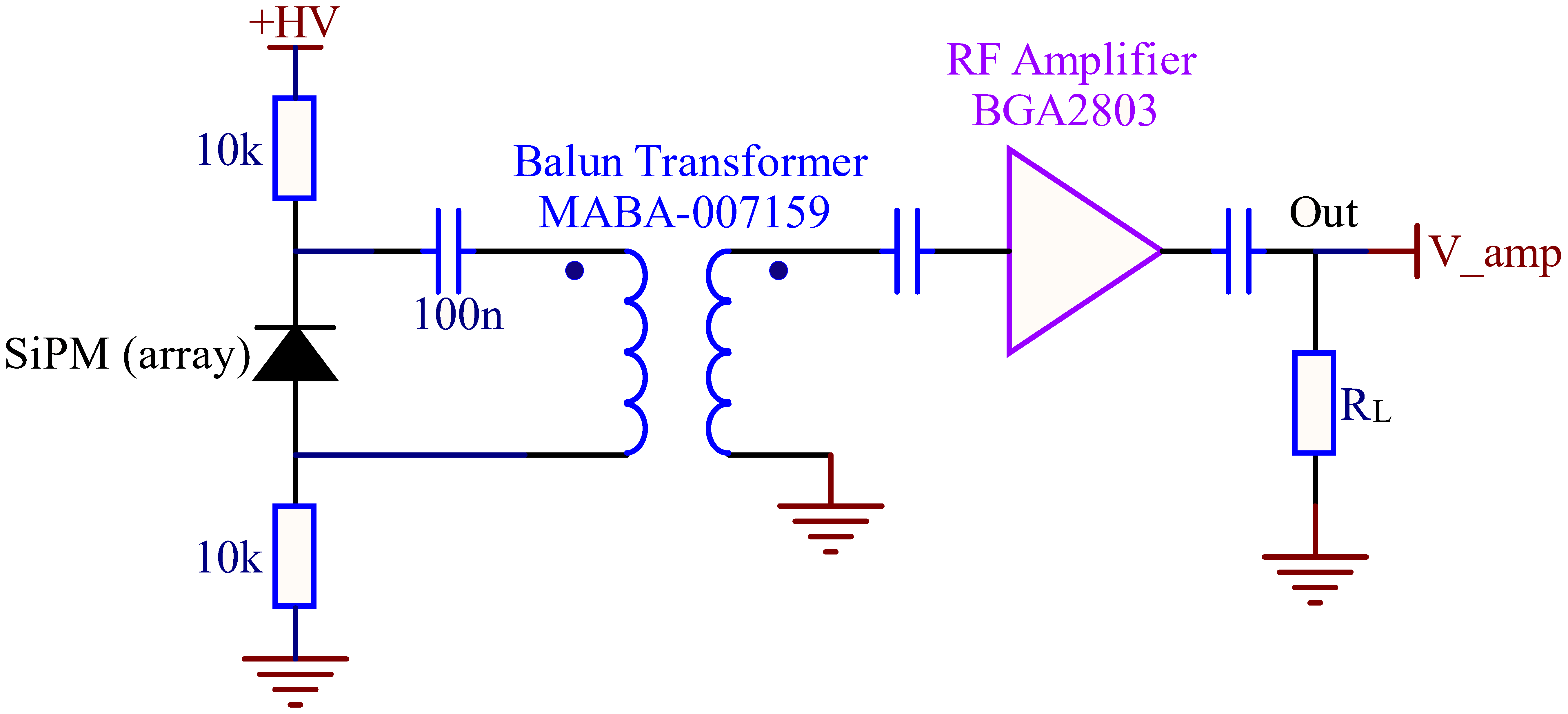}
    \caption{Schematic of the pre-amplifier circuit for SiPMs or SiPM arrays.
             These capacitors are utilized for AC coupling, such as isolating the DC level of the RF amplifier.}
    \label{fig:3.sipm_pre_amplifier}
\end{figure}

The readout of series-parallel connection SiPM arrays is achieved by a two-stage amplification, comprising a pre-amplifier and an analog summing circuit.
The primary requirement for amplification in the SiPM front-end readout is high bandwidth, as the signal from the SiPM (S13360-3050) has a rapid rise time of approximately 1 ns.
We design the pre-amplifier based on a BGA2803 RF amplifier, which not only offers a bandwidth of approximately 2 GHz, a gain of 23.5 dB, and a noise figure of 3.6 dB but also features low power consumption with each channel consuming about $17~\mathrm{mW}$~\cite{Low_power_RF_amplifiers, Datasheet_BGA2803}.
Besides, we utilize a balun transformer (MABA-007159) with a turn ratio of 1:1 and an impedance of $\mathrm{50~\Omega}$ to match the input impedance of the RF amplifier, as shown in figure~\ref{fig:3.sipm_pre_amplifier}.

\begin{figure}[htbp]
  \centering
  \includegraphics[width=0.8\textwidth]{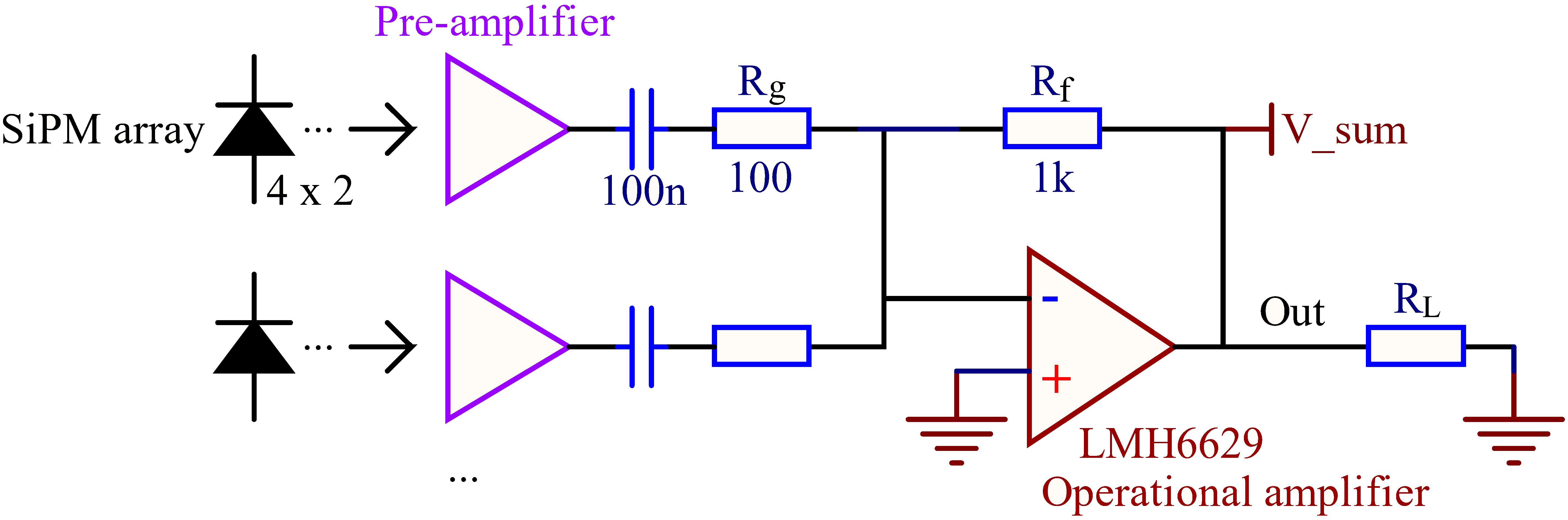}
  \caption{Schematic of multi-channel summing circuit design, which shows the summation of two series-parallel channels and additional channels can be appended in a similar way.}
  \label{fig:3.invert_summing_circuit}
\end{figure}

The summing circuit is designed based on an LMH6629 operational amplifier, which features a gain bandwidth product (GBP) of 4 GHz and a minimum stable gain of 10 V/V~\cite{Datasheet_LMH6629}.
As shown in figure~\ref{fig:3.invert_summing_circuit}, two series-parallel channels are summed to one readout channel, each combined by a pre-amplifier and a series-parallel combination SiPM array with four SiPMs in series and two series in parallel, resulting in a total of sixteen SiPMs.
To increase the phase margin and improve the loop stability, we set the gain of each input channel to $R_f/R_g=10$ V/V.
According to our simulation results, the bandwidth of the summing circuit in this configuration is about 270 MHz.
Additionally, the LMH6629 has a power consumption of about 56 mW, so the overall power consumption of the front-end readout for $4\times4$ SiPMs is about $17\times2+56=90$ mW.

The number of input channels is primarily limited by the decrease in SNR and bandwidth of the summing circuit.
For an input channel number of $N$, the baseline noise from different input channels will be summed to $\sqrt{N}$ of one channel, leading to a decrease of SNR by a factor of $1/\sqrt{N}$.
For instance, the SNR of the summing of four input series-parallel channels will be half that of one input channel.
Moreover, the bandwidth in this configuration is about $\frac{GBP}{N}\cdot\frac{R_\text{g}}{R_\text{f}}$ since the noise gain of the operational amplifier is $R_\text{f}/(R_\text{g}/N)$ for $N$ input channels.
This influence can be improved by reducing the gain of each channel.

\section{SPTR measurements}
\label{sec4}



\begin{figure}[htbp]
  \centering
  \begin{minipage}[c]{0.4\textwidth}
    \centering
    \includegraphics[width=1\textwidth]{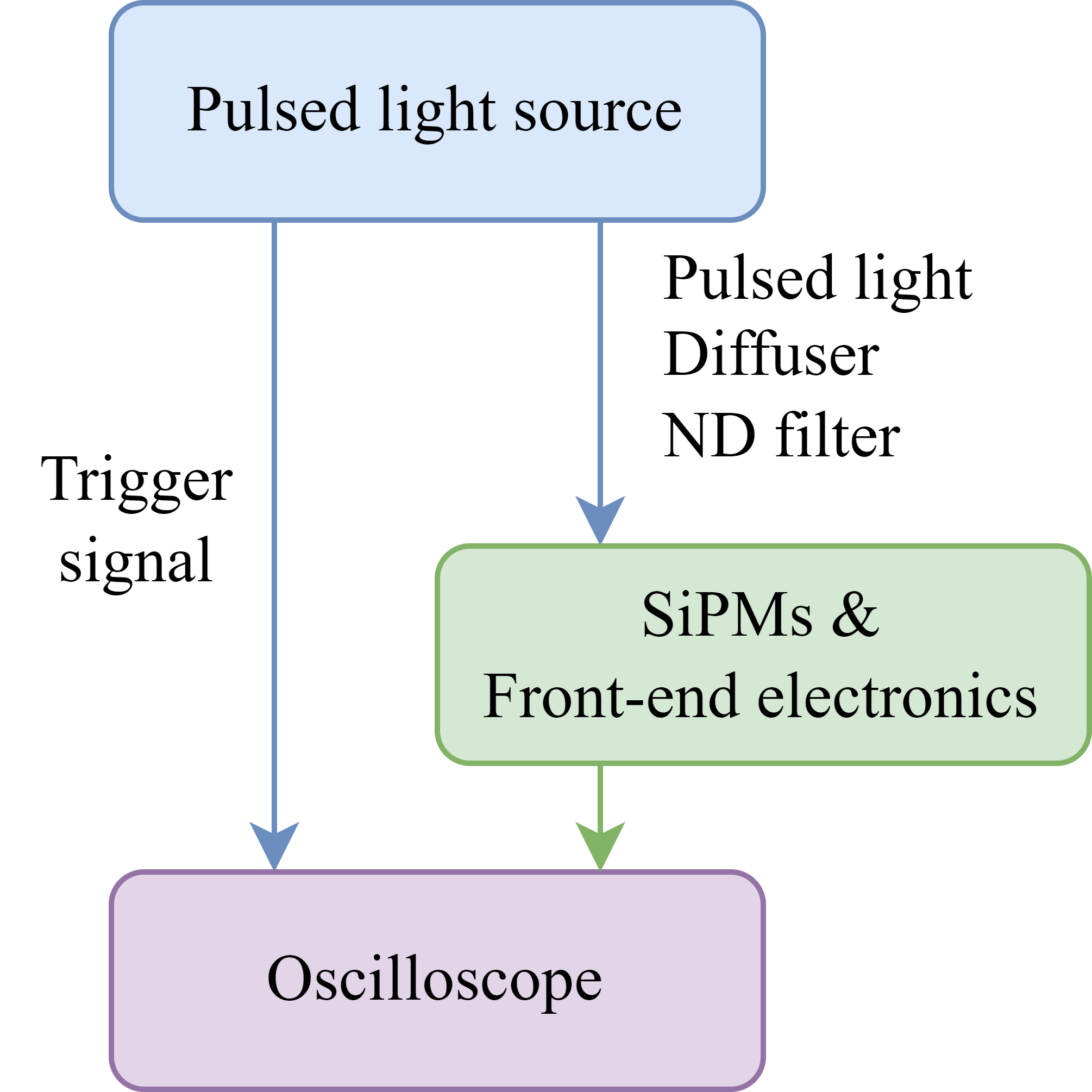}
  \end{minipage}
  \hfil
  \begin{minipage}[c]{0.4\textwidth}
    \centering
    \includegraphics[width=1\textwidth]{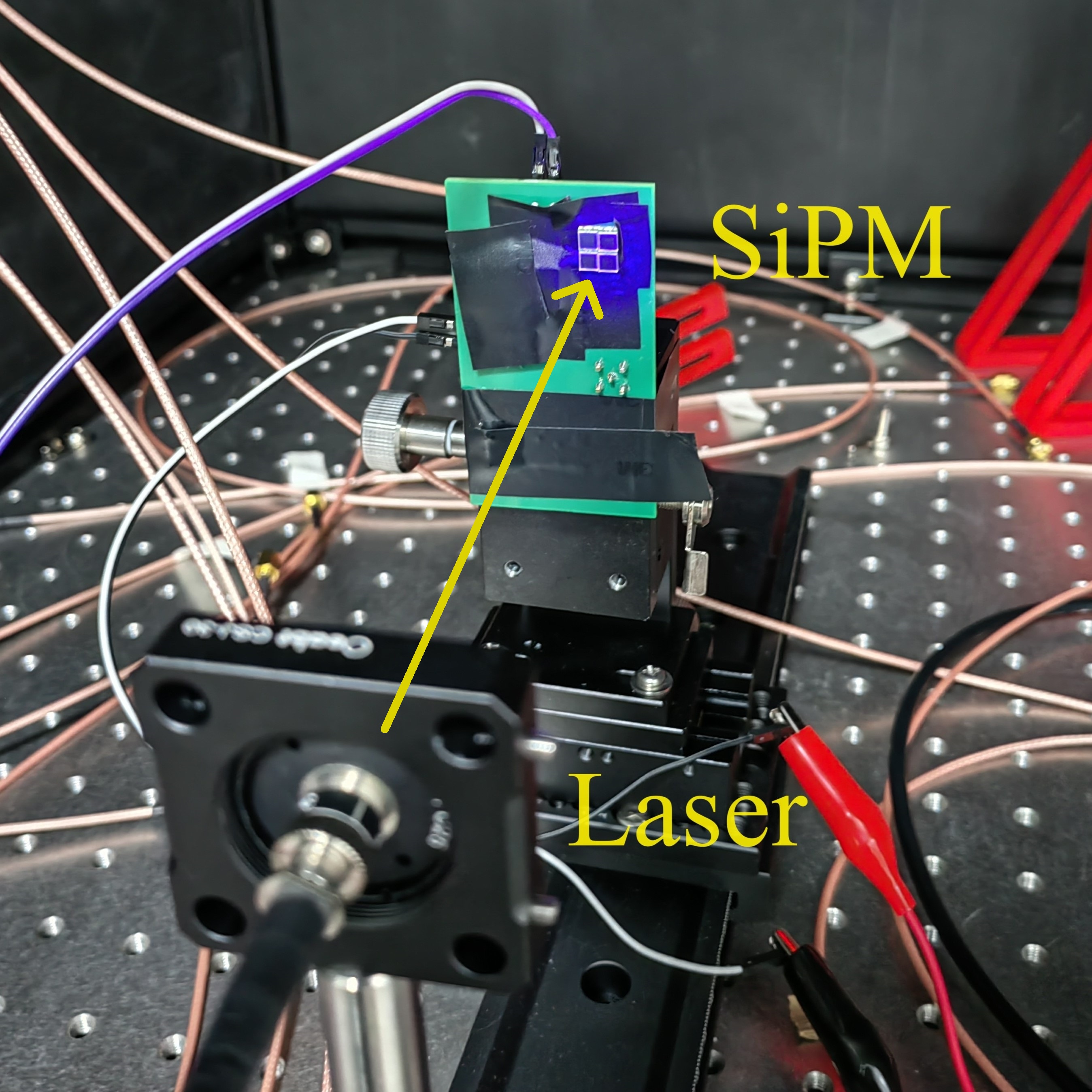}
  \end{minipage}
  \caption{Left: the flowchart of SPTR measurement.
           Right: a picture of the setup for SPTR measurement, where a $2\times2$ SiPM array with a configuration of two SiPMs in series and two series in parallel is utilized.}
  \label{fig:4.SPTR_experiment_setup}
\end{figure}

The test setup of SPTR measurement is illustrated on the left side of figure~\ref{fig:4.SPTR_experiment_setup}, primarily involving a narrow pulsed light source, SiPMs with their front-end readout electronics, and an oscilloscope.
The light source emits two signals.
One signal is the pulsed light that undergoes diffusion, attenuates to the single-photon level, and is ultimately received by SiPMs.
The signal from SiPMs is processed by the front-end readout electronics described in section~\ref{sec3} and is subsequently output to the oscilloscope for further data analysis.
The other signal is an electrical signal synchronized with the light signal, directly output to the oscilloscope as the reference of photons' arrival time.

A photo of the experiment setup is shown on the right side of figure~\ref{fig:4.SPTR_experiment_setup}.
The time duration of the pulsed light needs to be much smaller than the SPTR of SiPMs, thereby a 405 nm picosecond laser (Taiko PDL M1 LDH-IB-405-B) with a pulse width FWHM of less than 50 ps is used as the light source.
Moreover, diffusers and neutral density (ND) filters are utilized to diffuse and attenuate the light pulses.
The oscilloscope (TELEDYNE LECROY WavePro 254HD) provides a 20 GS/s sampling rate with 2.5 GHz bandwidth and 12-bit resolution.

\begin{figure}[htbp]
  \centering
  \includegraphics[width=0.7\textwidth]{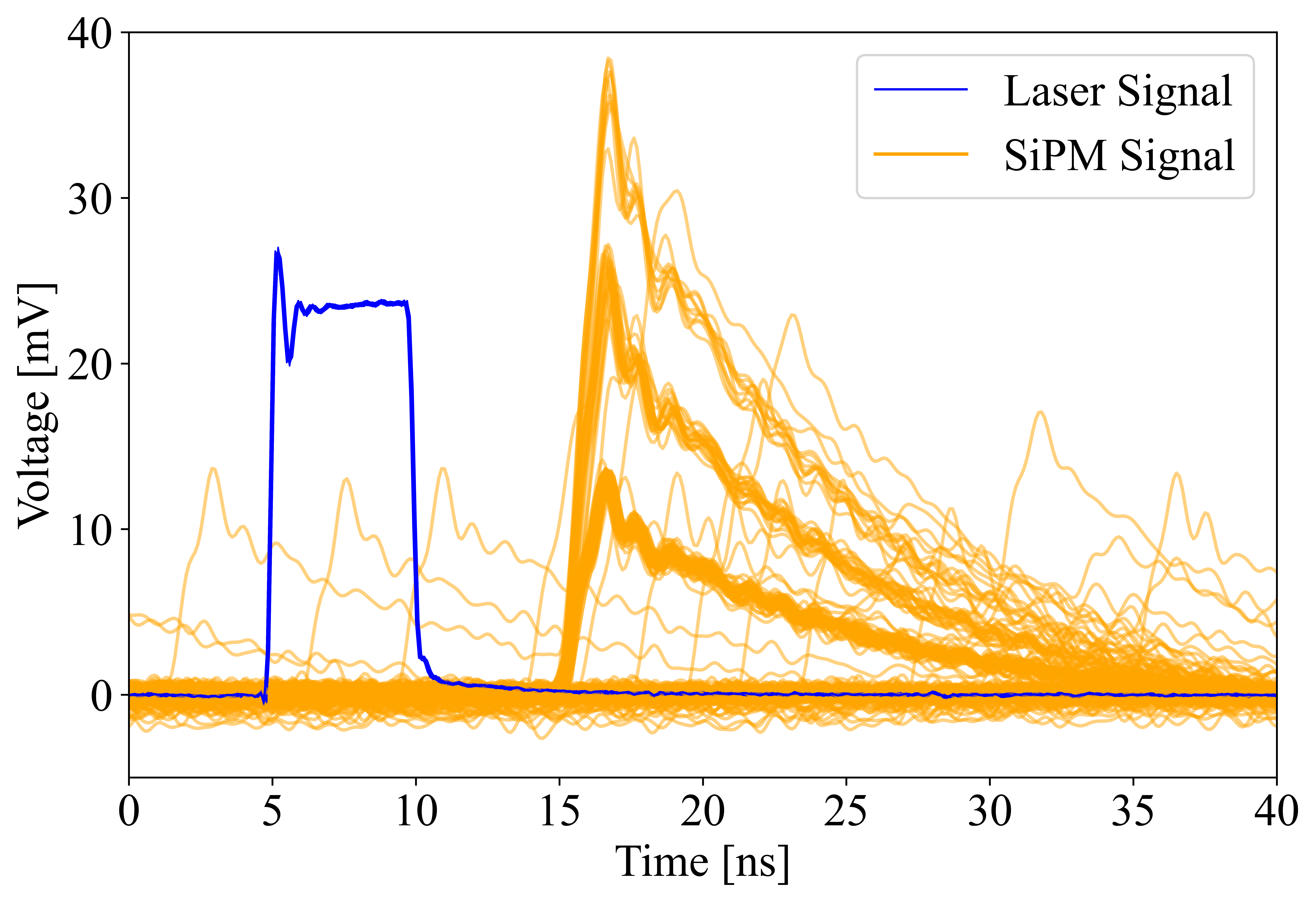}
  \caption{An example of 200 waveforms, where the SiPM is operated at 60 V bias voltage, and its signal is processed by the pre-amplifier.
  For ease of observation, the amplitude of the synchronized signal from the laser is scaled down to 1/40 and the timeline has been shifted.}
  \label{fig:4.single_waveform_sample}
\end{figure}
\begin{figure}[htbp]
  \centering
  \includegraphics[width=0.65\textwidth]{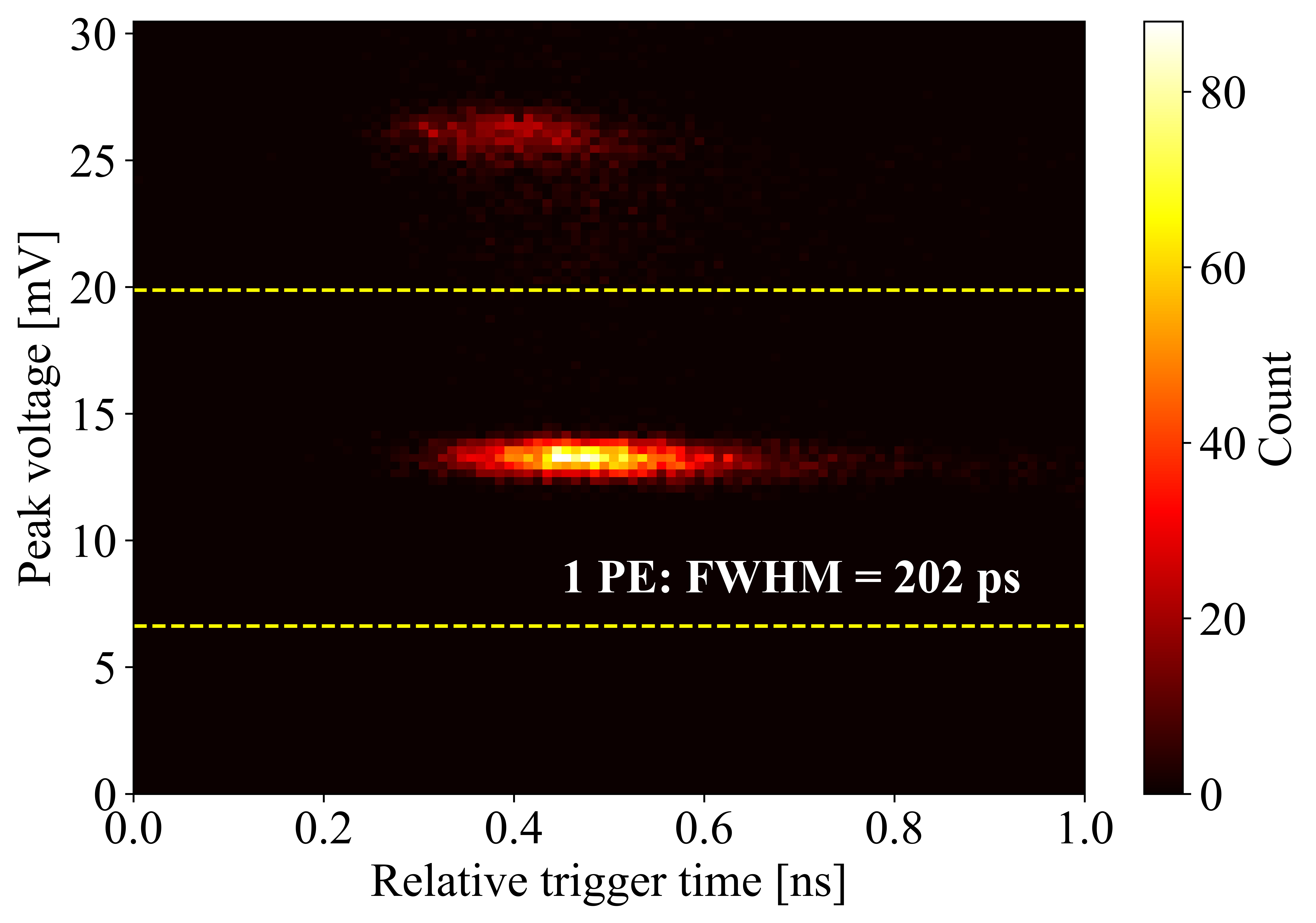}
  \caption{The distribution of the relative trigger time and the signal amplitude of a SiPM operating at 60 V bias voltage.
  The relative trigger time is the relative delay from when the laser signal is triggered, namely when the leading edge reaches the threshold, to when the SiPM signal is triggered.}
  \label{fig:4.single_result_sample}
\end{figure}

The timing method used for data analysis involves fixed threshold triggering of the leading edge, which is consistent with the timing approach of hDOM, i.e., triggering using high-speed comparators.
An example of waveforms for the pre-amplifier output from a SiPM is illustrated in figure~\ref{fig:4.single_waveform_sample}, and it is easy to distinguish the baseline, single-photon, and double-photon events.
The trigger time is searched within a pre-selected relatively narrow time window, such as from 10 ns to 20 ns in this figure, to minimize the ratio of dark noise events.
Additionally, linear interpolation is employed here to obtain a more refined result using the upper and lower points around the threshold, reducing the impact of the limited sampling rate of the oscilloscope.
As a result, the distribution of the relative trigger time and the signal amplitude is presented in the figure~\ref{fig:4.single_result_sample}, where the threshold voltage is 1.5 mV and the photon number is represented by the signal amplitude.
Applying a Gaussian fitting to the relative trigger time distribution within the single-photon region, where the peak voltage falls within the range of $\left[\frac{1}{2}V_{\text{1PE}}, \frac{3}{2}V_{\text{1PE}}\right]$, a SPTR of approximately 200 ps FWHM can be obtained.

\begin{figure}[htbp]
  \centering
  \includegraphics[width=0.7\textwidth]{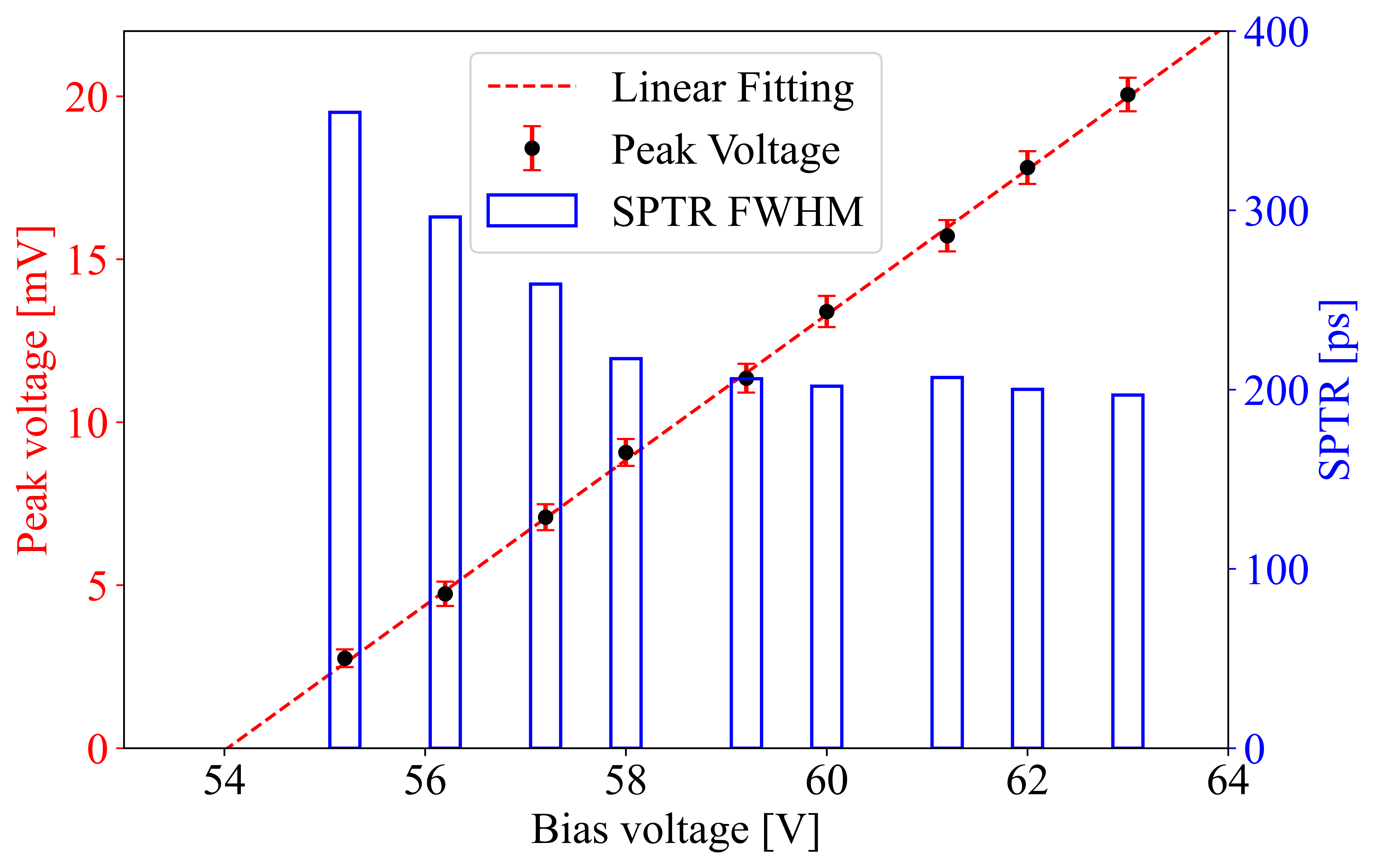}
  \caption{The signal amplitudes and SPTRs of a SiPM operating at different bias voltages.
           The breakdown voltage can be found to be about 54 V by linear fitting.}
  \label{fig:4.Vbias_scan_sample}
\end{figure}

The performance of the SiPM is closely related to the overvoltage, where a higher overvoltage leads to a higher gain but also results in a higher DCR and crosstalk rate, and more afterpulses.
To investigate the impact of overvoltage on SPTR and determine the appropriate operating voltage, we test the SPTR of SiPMs at different operating voltages.
The test result is shown in figure~\ref{fig:4.Vbias_scan_sample}, where the breakdown voltage can be obtained by linear fitting using the signal amplitudes at different bias voltages.
Since the SPTR measurement results vary at different threshold voltages, the best SPTR results are obtained by scanning the threshold voltages.
It can be observed from this graph that the SPTR measurement results stabilize at around 200 ps when the overvoltage exceeds 5 V.

\begin{figure}[htbp]
  \centering
  \begin{minipage}[c]{0.45\textwidth}
    \centering
    \includegraphics[width=1\textwidth]{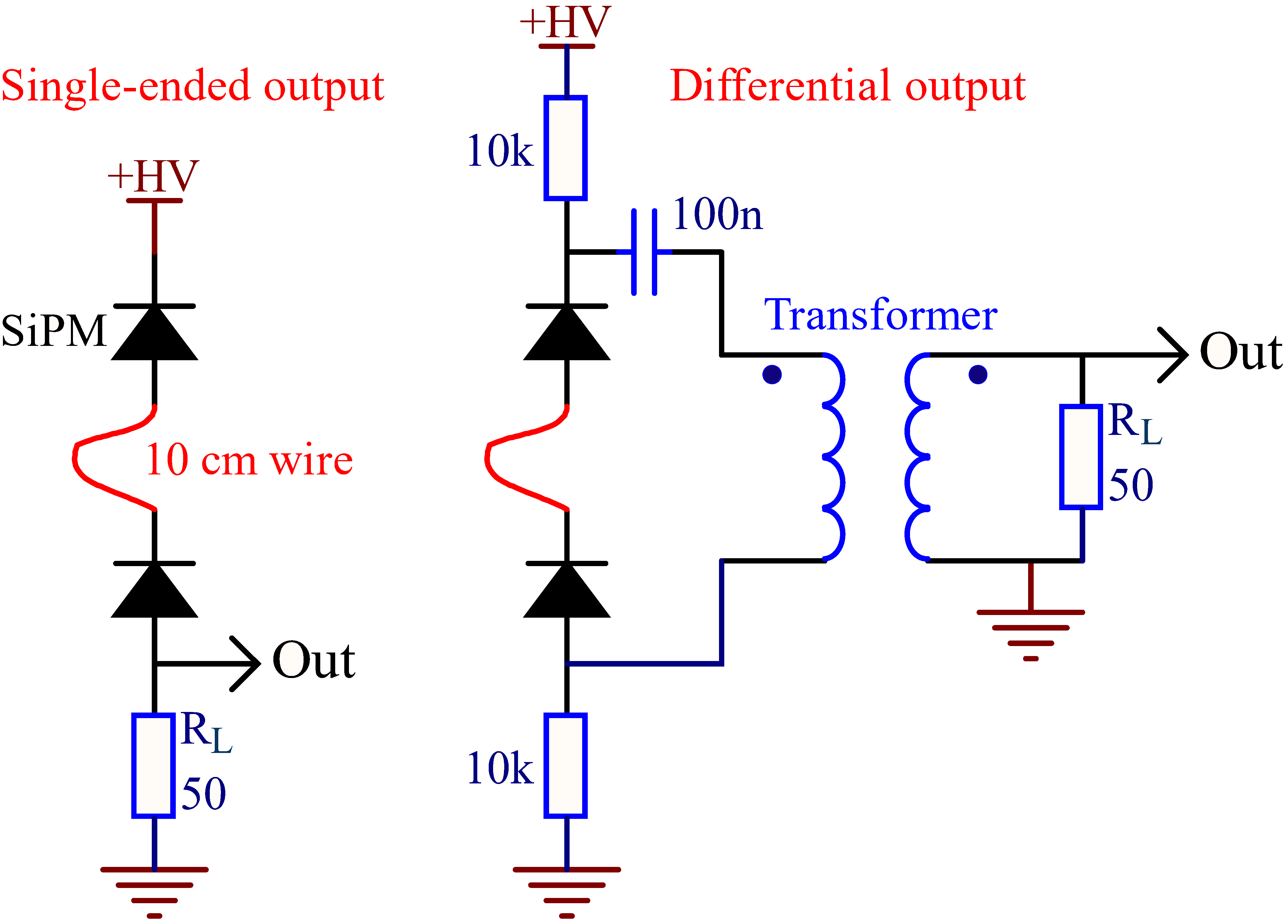}
  \end{minipage}
  \hfil
  \begin{minipage}[c]{0.5\textwidth}
    \centering
    \includegraphics[width=1\textwidth]{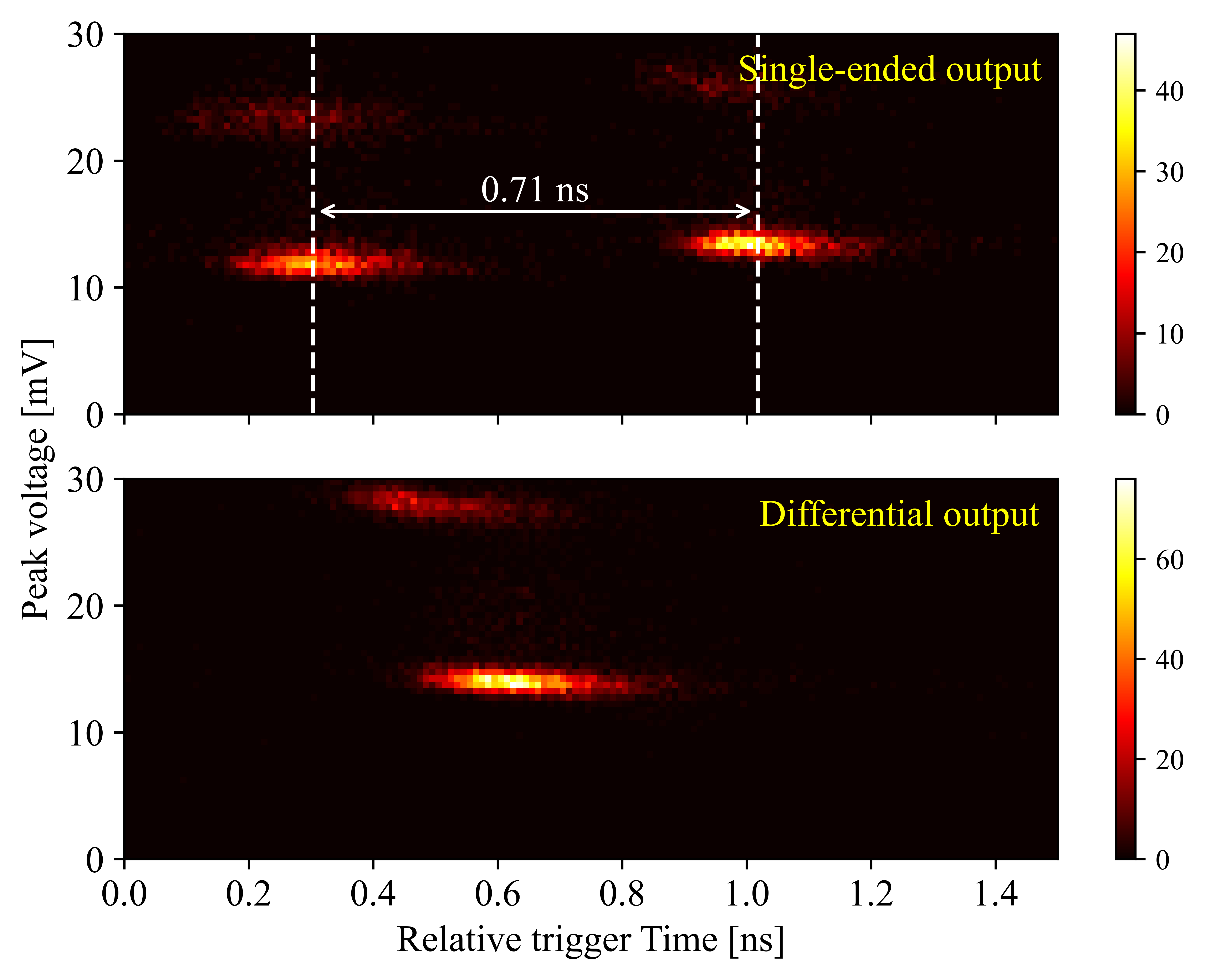}
  \end{minipage}
  \caption{Left: two schematics depicting two SiPMs connected in series using different output patterns: one with a single-ended output and the other with a differential output.
           The signal line between the two SiPMs spans about 10 cm for both two configurations.
           Right: the relative trigger time distributions of the two configurations.}
  \label{fig:4.trigger_times_2s}
\end{figure}

To demonstrate the improvement of signal path differences by differential output, we conducted experiments of a series-parallel SiPM array with two SiPMs in series, as shown in figure~\ref{fig:4.trigger_times_2s}.
We extended the signal line length between the two SiPMs in series to about 10 cm, resulting in an expected time delay of approximately 0.65 ns for the single-ended output signals of the two SiPMs.
The experiment result of single-ended output shows two distinct peaks for the single photon response in the relative trigger time, approximately 0.71 ns apart, consistent with expectations.
In contrast, the differential output result exhibits only one peak, indicating that the time delay caused by the $\text{10 cm}$ wire is alleviated.

\begin{figure}[htbp]
  \centering
  \begin{minipage}[c]{0.495\textwidth}
    \centering
    \includegraphics[width=1\textwidth]{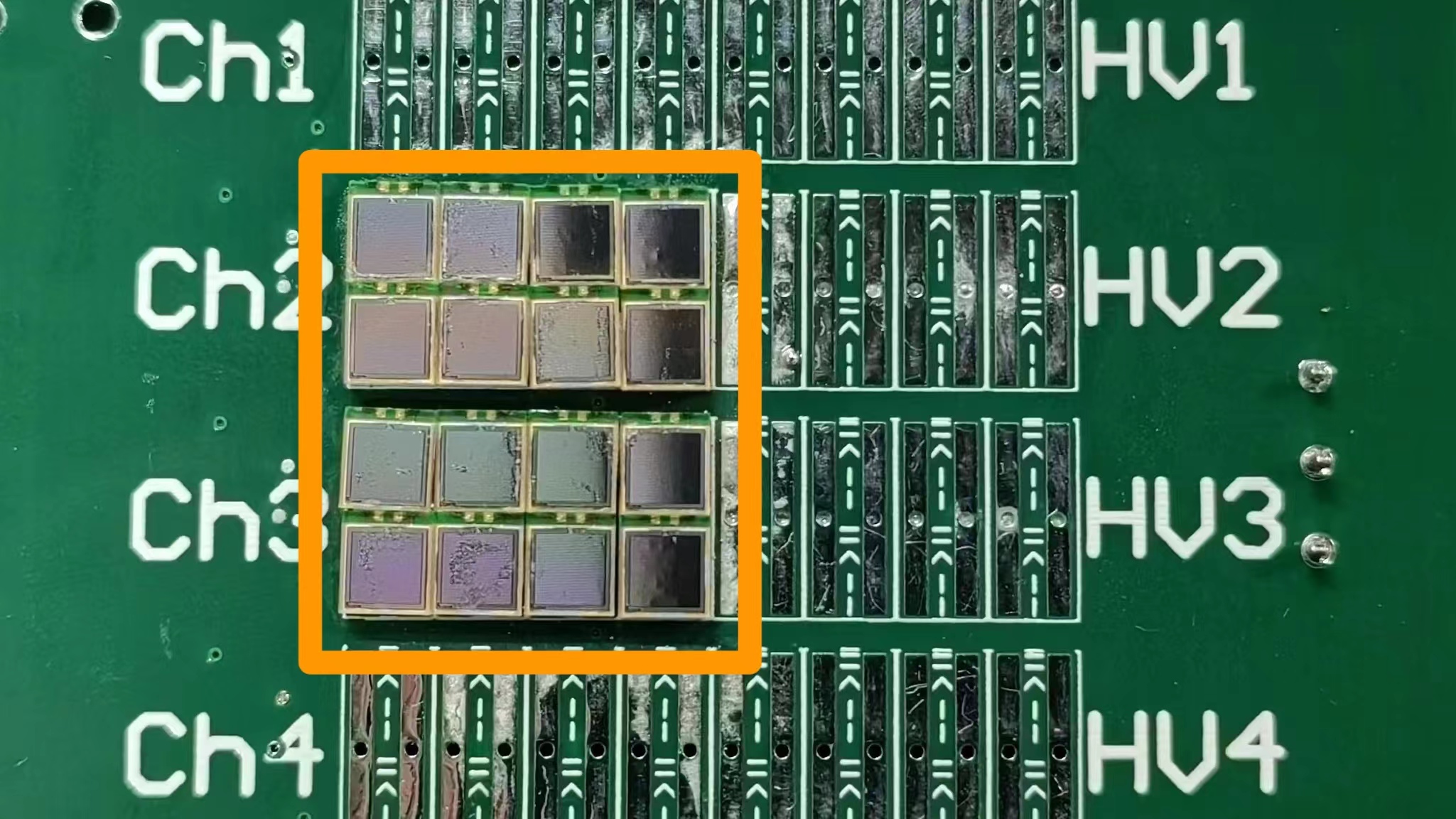}
  \end{minipage}
  \hfil
  \begin{minipage}[c]{0.495\textwidth}
    \centering
    \includegraphics[width=1\textwidth]{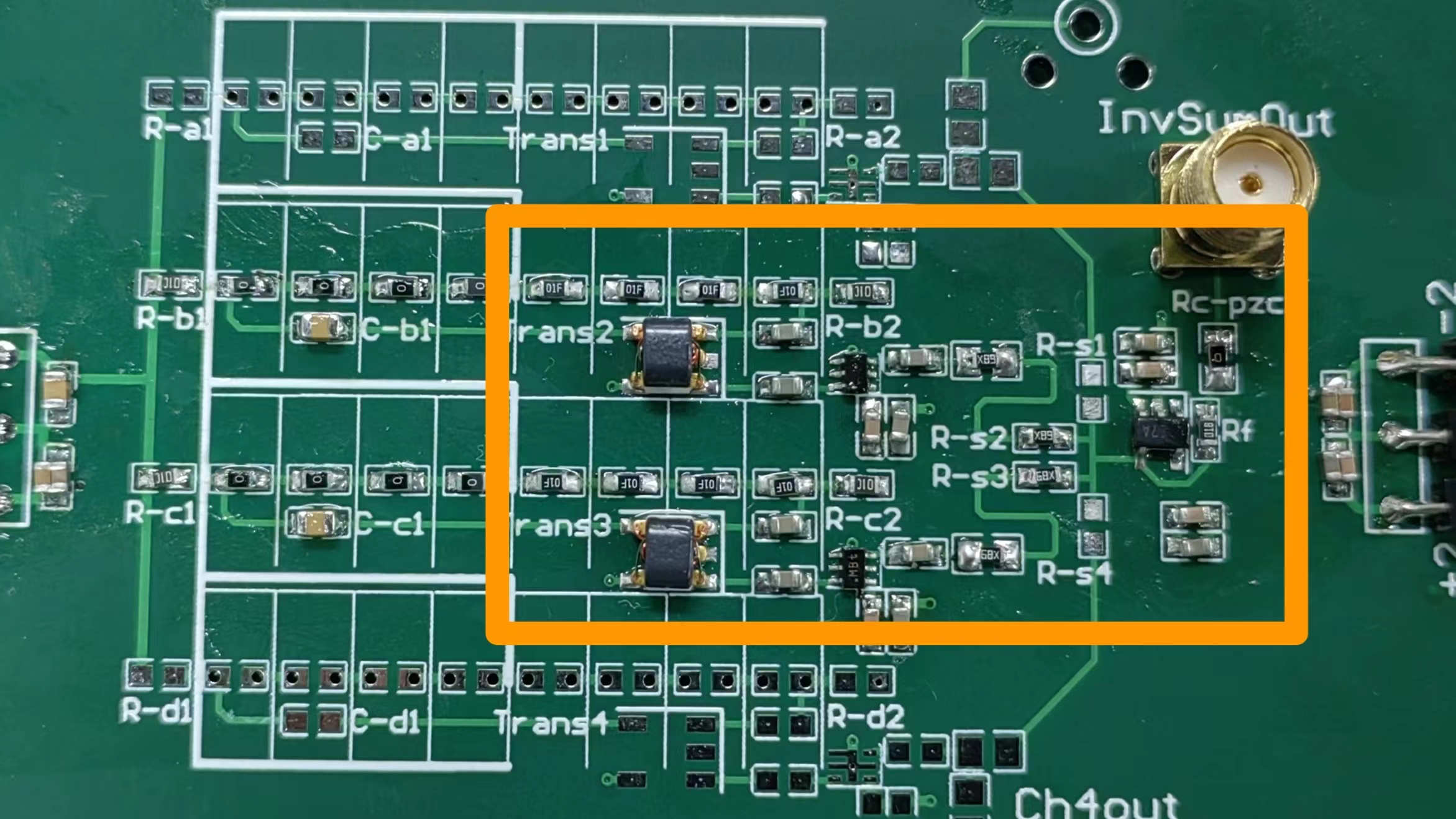}
  \end{minipage}
  \caption{Photos of the $4\times4$ SiPM array.
           This board is used to test the performance of SiPM arrays in different configurations, so there are some extra pads.}
  \label{fig:4.photo_array44}
\end{figure}

We tested the SPTRs of the SiPMs under three configurations, as shown in table~\ref{tab:SPTR_results}.
According to the results in figure~\ref{fig:4.Vbias_scan_sample}, each SiPM is operated at 60 V bias voltage to reduce the differences in the time performance among different SiPMs and alleviate the effects of time walk described in section~\ref{sec3}.
The $2\times2$ SiPM array is constructed with a series-parallel combination SiPM array, with two SiPMs in series and two series in parallel, along with a pre-amplifier.
The power consumption of the $2\times2$ SiPM array is similar to that of a single piece because they are both equipped with a pre-amplifier, which is the main source of power consumption.
Moreover, the $4\times4$ SiPM array is a summing output of two series-parallel channels, with each series-parallel array configured as four SiPMs in series and two series in parallel.
Two photos of the $4\times4$ SiPM array are shown in figure~\ref{fig:4.photo_array44}.
The operating voltages of the $2\times2$ SiPM array and $4\times4$ SiPM array are $120~\mathrm{V}$ and $240~\mathrm{V}$, respectively.
For the $4\times4$ SiPM array that has a detection area of $12\times12~\mathrm{mm}^2$, we obtain a SPTR of about 300 ps with an overall power consumption of approximately $90~\mathrm{mW}$.

\begin{table}[htbp]
  \centering
  \caption{Results of SPTR measurements, with each SiPM operating at 60 V bias voltage.
           The capacitance of a SiPM (S13360-3050PE) is approximately 320 pF.}
  \smallskip
  \begin{tabular}{c|c|c|c}
    \hline
    SiPM quantity & Detection area & Power consumption & SPTR FWHM \\
    \hline
    A single piece          & $3\times3~\mathrm{mm}^2$   & 17 mW & $\approx$ 200 ps \\
    A $2\times2$ SiPM array & $6\times6~\mathrm{mm}^2$   & 17 mW & $\approx$ 220 ps \\
    A $4\times4$ SiPM array & $12\times12~\mathrm{mm}^2$ & 90 mW & $\approx$ 300 ps \\
    \hline
  \end{tabular}
  \label{tab:SPTR_results}
\end{table}


For the single piece of SiPM with a measured SPTR of 200 ps FWHM, the contribution of electronic noise to the SPTR is given by $\text{jitter}=\frac{\sigma_{\text{noise}}}{dV_{\text{th}}/dt}\approx50$ ps FWHM, which is not the main factor to SPTR.
This value for the $2\times2$ SiPM array increases to $\text{jitter}\approx 90$ ps FWHM due to the decrease of signal amplitudes, primarily caused by the parallel connection.
The jitter for the $4\times4$ SiPM array is further increased to approximately 170 ps FWHM, mainly because the baseline noise is amplified by the summing circuit.
Electronic noise is a significant factor affecting the time performance of large-area SiPM arrays.
One potential improvement strategy involves selecting SiPMs with larger micro-cell sizes to enhance the gain and SNR.


\section{Conclusions}

TRIDENT is exploring the improvement of angular resolution by employing SiPMs with superior time resolution performance in its detection units (hDOM).
In this article, we analyzed the challenges in the front-end readout of large-area SiPM arrays and explained that the primary challenge is achieving a high-precision SPTR while increasing the area of the SiPM array.
We designed a series-parallel connection SiPM array scheme and analyzed the possible influences of the series and parallel connections.
In addition, we designed a two-stage amplification scheme, consisting of a pre-amplifier based on a balun transformer and RF amplifier, and a multi-channel summing circuit.
Finally, we conducted tests on the SPTR of SiPMs and obtained the single photon time resolution of a $4\times4$ SiPM array ($12\times12~\mathrm{mm}^2$) of approximately 300 ps FWHM with a power consumption of about 90 mW.
We also discussed the primary factors influencing the time performance and proposed potential approaches for improvement.


\acknowledgments
We thank Jun Guo, Hualin Mei, Yong Yang, Wei Tian, Fuyudi Zhang, Jingtao Huang and Qichao Chang for their help to improve this paper.
This work was sponsored by the Ministry of Science and Technology of China (No. 2022YFA1605500), Shanghai Pilot Program for Basic Research — Shanghai Jiao Tong University (No. 21TQ1400218), Yangyang Development Fund, Office of Science and Technology, Shanghai Municipal Government (No. 22JC1410100), Shanghai Jiao Tong University under the Double First Class startup fund and the Foresight grants (No. 21X010202013) and (No. 21X010200816).


\bibliographystyle{JHEP}
\bibliography{main.bib}

\end{document}